\documentclass[aps,pre,reprint,floatfix,superscriptaddress]{revtex4-2}

\usepackage{graphicx}
\usepackage{hyperref}
\usepackage{amsmath,amssymb,amsfonts}
\usepackage{algpseudocode}

\begin{document}

\title{Random nanowire networks: Identification of a current-carrying subset of wires  using a modified wall follower algorithm}

\author{Yuri~Yu.~Tarasevich}
\email[Corresponding author: ]{tarasevich@asu.edu.ru}
\affiliation{Laboratory of Mathematical Modeling, Astrakhan State University, Astrakhan 414056, Russia}

\author{Renat~K.~Akhunzhanov}
\email{akhunzha@mail.ru}
\affiliation{Laboratory of Mathematical Modeling, Astrakhan State University, Astrakhan 414056, Russia}

\author{Andrei~V.~Eserkepov}
\email{dantealigjery49@gmail.com}
\affiliation{Laboratory of Mathematical Modeling, Astrakhan State University, Astrakhan 414056, Russia}

\author{Mikhail V. Ulyanov}
\email{muljanov@mail.ru}
\affiliation{V. A. Trapeznikov Institute of Control Sciences of RAS, Moscow 117997, Russia}
\affiliation{Computational Mathematics and Cybernetics, M. V. Lomonosov Moscow State University, Moscow 119991, Russia}

\date{\today}

\begin{abstract}
We mimic random nanowire networks by the homogeneous, isotropic, and random deposition of conductive zero-width sticks onto an insulating substrate. The number density (the number of objects per unit area of the surface) of these sticks is supposed to exceed the percolation threshold, i.e., the system under consideration is a conductor.  To identify any current-carrying part (the backbone) of the percolation cluster, we have proposed and implemented a modification of the well-known wall follower algorithm---one type of maze solving algorithm. The advantage of the modified algorithm is its identification of the whole backbone without visiting all the edges. The complexity of the algorithm depends significantly on the structure of the graph and varies from $O\left(\sqrt{N_\text{V}}\right)$ to $\Theta(N_\text{V})$. The algorithm has been applied to backbone identification in networks with different number densities of conducting sticks. We have found that (i) for number densities of sticks above the percolation threshold, the  strength of the percolation cluster quickly approaches unity as the number density of the sticks increases; (ii) simultaneously, the percolation cluster becomes identical to its backbone plus simplest dead ends, i.e., edges that are incident to vertices of degree 1. This behavior is consistent with the presented analytical evaluations.
\end{abstract}

\maketitle

\section{Introduction\label{sec:intro}}

A poorly conductive film containing randomly distributed highly conductive elongated fillers such as nanowires, nanotubes, and nanorods is an example of the random nanowire networks (NWNs)~\cite{Nam2016N,Ackermann2016SR,Callaghan2016PCCP,McCoul2016AEM,Zhang2017JMM,Hicks2018JAP,Glier2020,Balberg2020}. A possible application of random NWNs is in memristive elements and neuromorphic systems~\cite{OCallaghan2018,DiazAlvarez2019,DiazAlvarez2020,Li2020,Kuncic2021}.

In a two-component, random mixture, there is a critical concentration for a component when this component forms a connected subset (a percolation cluster) that spans the opposite boundaries of the disordered medium. Different kinds of percolation processes have attracted the attention of the scientific community for several decades~\cite{Kesten1982,Stauffer,Sahimi1994,Grimmett1999,BollobasRiordan2006}. The physical properties of the disordered medium drastically change when a percolation cluster occurs. For instance, when the disordered medium consists of conductive and insulating substances, an insulator--conductor phase transition can be observed as the concentrations of the two components vary. The relation between the electrical conductivity behavior and percolation as a phase transition can be found in Ref.~\onlinecite{Balberg2020}. However, only a part of the percolation cluster carries the electrical current~\cite{Skal1975SPS,DeGennes1976}. The set of current-carrying edges of the  percolation cluster is called the (effective) backbone~\cite{Herrmann1984JPhA}. The rest of the percolation cluster is a set of dead ends~\cite{Herrmann1984JPhA} (also called dangling ends~\cite{Porto1999}, tag ends~\cite{Kirkpatrick1978AIPCP}, tangling ends~\cite{Grassberger1992JPhA}) and perfectly balanced bonds (Wheatstone bridges). Since the potential difference between the ends of a perfectly balanced bond is equal to zero, electrical current through this bond is absent~\cite{Li2007JPhA}.

Special attention has been paid to percolation in disordered systems produced by the random deposition of elongated particles onto a substrate~\cite{Li2009PRE,Mertens2012PRE,Li2013PRE,Li2016PhysA,Lin201917PT,Tarasevich2020PRE}. To mimic the real-world elongated particles and, at the same time, simplify the simulations, line segments (sticks)~\cite{Li2009PRE,Mertens2012PRE} and different simple geometric shapes are used, e.g., rectangles~\cite{Li2013PRE}, ellipses~\cite{Li2016PhysA}, superellipses~\cite{Lin201917PT}, and discorectangles.\cite{Tarasevich2020PRE} With increase of the aspect ratio, $\varepsilon$, each of these shapes tends to a zero-width stick ($\varepsilon \to \infty$). For zero-width sticks of equal length randomly oriented and placed onto a plane, the best encountered value of the percolation threshold is
$n_\text{c}^\times = 5.637\,285\,8(6)$
(see Ref.~\onlinecite{Mertens2012PRE}). A synopsis of the experimental data~\cite{Vodolazskaya2019JAP} evidences that the typical aspect ratio for Ag nanowires ranges  from 100 to 1000. For all the above listed shapes of such large aspect ratio, the percolation threshold coincides with $n_\text{c}^\times$ to within few percents. Curved~\cite{Hicks2018JAP,Lee2021} and wavy~\cite{Berhan2007,Li2008} lines are also used in simulations to better mimic the shape of wires, since real-world nanowires are not perfectly straight.

A calculation has been presented for the excluded area between penetrable rectangles in 2D as a function of the aspect ratio and orientational order parameter~\cite{Chatterjee2015JSP}.
For isotropically distributed systems, the percolation thresholds for different values of the aspect ratio are in close agreement with the findings from Monte Carlo simulations~\cite{Li2013PRE}.

In a random graph (e.g., in a percolation cluster), the geometrical backbone is defined as a union of all the self-avoiding walks (SAWs) between the two given vertices of this graph~\cite{Shlifer1979JPhysA}. A SAW or a simple path is a path that contains no vertex twice. One  algorithm for finding simple paths in a graph is based on depth-first search~\cite{Hopcroft1973CACM}. The geometrical backbone consists of the effective backbone and of the perfectly balanced bonds. In other words, the effective backbone is the set of all edges that carry a current, while the geometrical backbone is the complete set of both current-carrying edges and perfectly balanced edges~\cite{Batrouni1988PRA}.

So-called red bonds or singly connected bonds are those that carry the total current; when they are cut, the current flow stops~\cite{Bunde1991percolationI}. Such red bonds cause locally inhomogeneous heating of nanowire-based networks (these so-called hotspots in the samples  accelerate nanowire degradation, leading to electrode failure)~\cite{Sannicolo2016NanoLett,Kumar2017JAP,Khaligh2017,Sannicolo2018,Kim2021}.

Potentials and currents in any random resistor network (RRN) can be found using Ohm's law and Kirchhoff's rules~\cite{Kirkpatrick1971PRL,Kirkpatrick1973RMP,Li2007JPhA,Benda2019,Kim2020JCPC}. Thus, in principle, the current-carrying subset of the whole RRN may be extracted. However, since calculations of electrical potentials and currents are based on floating-point arithmetic, round-off errors are unavoidable.  Due to these round-off errors, some false apparent currents may arise both in dead ends and in perfectly balanced bonds. These currents impede the correct backbone extraction. Moreover, direct calculations of potentials and currents within an RRN deal with huge systems of linear equations and require a lot of computer memory. Only relative small systems can be treated in these approaches because the number of equations to be solved is proportional to the square of the linear size of the system under consideration. Hence, prior extraction of the geometrical backbone may significantly reduce the computations of currents and potentials by excluding senseless attempts to calculate currents in the dead ends. The improvement is expected to be most significant just above the percolation threshold, when dead ends are a dominant part of the percolation cluster.

To extract a geometrical backbone, search algorithms on graphs can be applied~\cite{Tarjan1972SIAM,Roux1987JPhA,Moukarzel1998IJMPhC,Herrmann1984JPhA,Herrmann1984PRL,Grassberger1992JPhA,Mastorakos1993PRE,Porto1997PRE,Babalievski1998IJMPC,Sheppard1999JPhysA,Yin2000PhysB,Yin2003IJMPC,Trobec2017AES}.
In fact, some of the algorithms for backbone identification belong to maze solving algorithms (such as the ``Ariadne's clew algorithm''~\cite{Mazer1998ACA}), which, in particular, are also applied to wire routing on chips~\cite{Fattah2015NOCS}. Some of these algorithms require storing, not only the original network, but its dual~\cite{Roux1987JPhA}. Stack overflow should be kept in mind when the backbone identification is based on the breadth-first search and depth-first search algorithms. All of the available graph-based algorithms remain storage limited, as some information at each node of the graph remains necessary~\cite{Alava2001}. In fact, application of these algorithms is also restricted to RRNs of moderate size. A comparison and analysis of the algorithms devoted to identification of the current-carrying part of the RRN demonstrates both the advantages and disadvantages of these above approaches~\cite{Tarasevich2018JPhCSbackbone}.

The goal of the present work is an investigation of the electrical properties and critical behavior of 2D disordered systems with an insulating host matrix and conductive rod-like fillers (zero-width sticks). The number density of fillers, $n$, ranges from the percolation threshold $n_\text{c}$ to  $n \approx 2.7 n_\text{c}$.

The rest of the paper is constructed as follows. Section~\ref{sec:methods} describes some technical details of the simulation and our modification of the wall follower algorithm that extracts the  geometrical backbones of the percolation clusters. Section~\ref{sec:results} presents our main findings. Section~\ref{sec:concl} summarizes the main results.

\section{Methods\label{sec:methods}}

\subsection{Modified wall follower algorithm}\label{subsec:MWFA}
Let $G$ be a connected plane graph, i.e., a particular embedding of a planar graph into a plane. We are looking for a set of all the SAWs between the two given vertices (entry, $V_\text{in}$, and exit, $V_\text{out}$) belonging to the outer perimeter of $G$. Let the graph $G$ be treated as a maze, i.e., its edges are considered as passages, while its vertices correspond to crossroads ($\deg V >1$) and dead ends ($\deg V =1$). In this case, one SAW may obviously be found by applying a wall follower algorithm. A walker proceeds from the entry until it reaches the exit while keeping  its left hand in contact with the wall of the maze. No passage (edge) can belong to a SAW if it is passed twice in opposite directions. Each twice traversed crossroads (vertex $\deg V >1$) indicates a simple cycle; any simple cycle cannot belong to a SAW and, hence, should be cut off. In such a way, the walker can find the leftmost SAW from the entry to the exit. Let this SAW be denoted as SAW0. Obviously, SAW0 is a part of the geometrical backbone. When $V_i$ and $V_j$ are the two distinct vertices belonging to SAW0, any SAW between $V_i$ and $V_j$ is also a part of the geometrical backbone. Moreover, any part of a graph enclosed  by a cycle cannot be a part of its backbone if fewer than two vertices of this cycle belong to the backbone.

In this way, the geometrical backbone can be found by using these two steps.
\begin{description}
  \item[Step 1] Looking for a SAW belonging to the backbone:
  \begin{itemize}
    \item Using the left-hand rule, looking for a SAW between $V_\text{in}$ and $V_\text{out}$;
    \item Push into a stack all the vertices of this SAW from $V_\text{out}$ to $V_\text{in}$;
    \item Mark the SAW as a part of the backbone.
  \end{itemize}
  \item[Step 2] While the stack is not empty, look for other SAWs belonging to the backbone:
    \begin{itemize}
  \item For each vertex in the stack, look for a SAW between this vertex and another vertex belonging to the already found part of the backbone.
      \item Push in the stack each vertex of any newly found SAW.
      \item Pop from the stack each vertex that has no incident untraversed edges.
      \end{itemize}
\end{description}
Some additional technical tricks may be useful to simplify and unify the search procedure~\cite{Akhunzhanov2021}.

While the algorithms searching for the backbone of a percolation cluster are mainly focused on lattice models of percolation theory, e.g.,~\cite{Herrmann1984JPhA,Roux1987JPhA,Grassberger1992JPhA,Trobec2017AES}, our algorithm can be used for both lattice and off-lattice problems; universal algorithms (for example, \cite{Tarjan1972SIAM}) can hardly take into account the specifics of the percolation cluster produced by the random deposition of rods onto a plane (a particular embedding of a planar graph into a plane).

Bearing in mind random NWNs, some particular kinds of graphs may be constructed as limiting cases, allowing complexity estimation of the algorithm.
\paragraph{Worst-case complexity estimate.}
The worst case is obviously when all edges of the graph $G$ belong to its backbone. For example, we can consider a square lattice $L \times L$ vertices. For this square lattice, the total number of vertices is $N_\text{V} = L ^ 2$, while the total number of its edges is $N_\text{E} =2L (L-1)$, which also estimates as $\Theta(N_\text{V})$. During the execution of the algorithm, each vertex is pushed in the stack and processed, while each edge is traversed twice (first as initially  traversed, and then as belonging to the backbone). Omitting the lower-order asymptotic terms, we obtain the worst-case complexity as $\Theta(N_\text{V})$.

\begin{figure}[!htb]
\includegraphics[width=0.65\columnwidth]{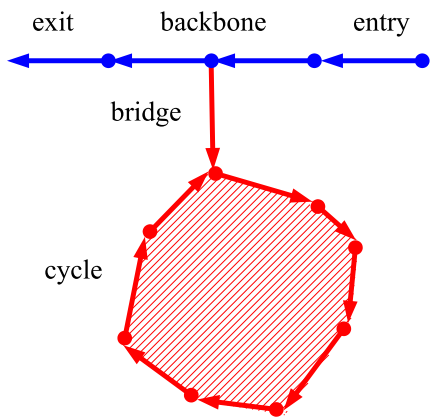}\\
\caption{Almost the whole graph is enclosed by a cycle; only one linear SAW connects the entry and exit.\label{fig:bestcase}}
\end{figure}

\paragraph{Best-case complexity estimate.}
The best case may be described as follows. A backbone is the one shortest linear SAW between the entry and exit, while the rest of the graph is enclosed by a cycle connected to the backbone by an articulation point or by a bridge (Fig.~\ref{fig:bestcase}). To identify the backbone, only the edges and vertices that belong to the outer perimeter of the graph are sufficient for the processing. The fraction of the edges and vertices belonging to the external perimeter of a graph depends significantly on the structure of the graph. However, this number of vertices hardly   exceeds $\sqrt{N_\text{V}}$ except for in some purposely constructed graphs. Hence, the complexity may be estimate as $O\left(\sqrt{N_\text{V}}\right)$.

These two limiting cases provide reasonable estimate range. The best-case consideration is closely related to incipient percolation cluster, while the worst-case consideration may be treated as corresponding to dense systems. Obviously, in the general case, the running time of the algorithm will be in between the estimates for the two extreme cases, approaching one or the other limiting estimate, depending on the structure of the system under consideration.

The red bonds of the percolation cluster can be identified as follows. Let us find the leftmost SAW and the rightmost one between $V_\text{in}$ and $V_\text{out}$. These SAWs can be found using the left-hand rule starting by moving once from $V_\text{in}$ to $V_\text{out}$ and then from $V_\text{out}$ to $V_\text{in}$. The edges that belong simultaneously to the both SAWs, if any, are the red bonds. See Supplemental Material at [URL will be inserted by publisher] for an animation which illustrates identification of red bonds in one particular percolation cluster.

\subsection{Application of the modified wall follower algorithm to a particular system}\label{subsec:MWFAapll}
A particular case of a plane graph involves the situation with $N$ zero-width sticks of length $l$, the centers of which are assumed to be independent and identically distributed (i.i.d.)  within a square domain $\mathcal{D}$ of size $L \times L$ with periodic boundary conditions; $\mathcal{D} \in \mathbb{R}^2$, i.e., $x,y \in [0;L]$, where $(x,y)$ are the coordinates of the center of the stick under consideration. Their orientations are assumed to be equiprobable. Hence, a homogeneous and isotropic network is produced. The relation $L>l$ is assumed. In our simulations, without loss of generality, sticks of unit length were used ($l = 1$). These sticks were randomly deposited onto $\mathcal{D}$  until the desired number density was reached. For basic computations, we used a  system of size $L=32$. The finite-size effect has additionally been tested by means of system size variation.

Each stick was treated as a resistor with a specified electrical conductivity, $\sigma$, i.e., an  RRN was considered. When a random graph is obtained using i.i.d. zero-width sticks, the degree (valence) of any of its vertices may be only 1 or 4. Vertices of degree 1 correspond to stick ends, while vertices of degree 4 correspond to intersections of the two sticks. Let a stick be intersected by $i$ other sticks. These intersections divide the stick into $i+1$ segments. Obviously, any segments, that is incident to a vertex of degree 1, is a dead end, hence, it cannot contribute to the electrical conductivity. A fraction of segments, which could potentially carry a current, is
\begin{equation}\label{eq:betaTDLn}
  P_\text{b} = 1 - \frac{\pi}{ n l^2} + \left( 1 + \frac{\pi }{n l^2}\right)\exp\left(-\frac{2 n l^2}{\pi}\right).
\end{equation}
Formula~\eqref{eq:betaTDLn} offers a theoretical estimate of the approximate backbone. A derivation of the formula~\eqref{eq:betaTDLn} can be found in Ref.~\onlinecite{Kim2018JAP}. It has also been obtained using some simplifying assumptions, including the Poisson distribution instead of the binomial one~\cite{Kumar2017JAP}.

\begin{figure}[!htb]
\centering
\includegraphics[width=\columnwidth]{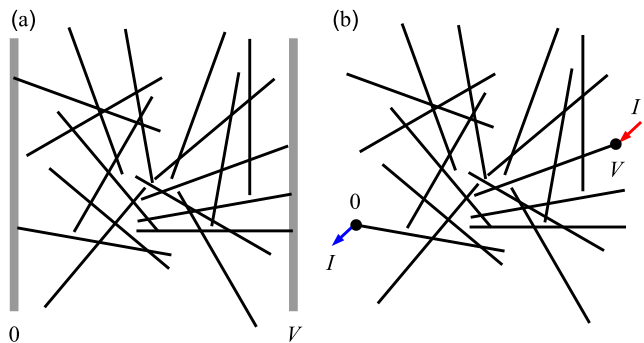}
\caption{\label{fig:geometry} Example of a random network produced by random isotropic deposition of equally-sized sticks. Each deposited stick is associated with a given conductance. (a)~bus-bar geometry; the (super)conducting buses are shown in gray. (b)~two-point geometry; source and sink sites are shown as closed circles.}
\end{figure}

When this network is a subject to a potential difference (say, 0 and $V$, where $V > 0$), there are two natural possibilities~\cite{Redner2009}, viz.,
\begin{itemize}
  \item the ``bus-bar geometry'', when two parallel (super) conducting bars (buses) are attached to the opposite borders of the network; and a potential difference is applied between these buses~\cite{Roux1987JPhA,Grassberger1992JPhA,Trobec2017AES} [Fig.~\ref{fig:geometry}(a)],
  \item the ``two-point geometry'', when a potential difference is applied between two distinct sites, so that an electrical current, $I$, injected into one site (source) and the same current is withdrawn from the other (sink)~\cite{Herrmann1984JPhA} [Fig.~\ref{fig:geometry}(b)].
\end{itemize}
In the case of superconducting buses, the ``bus-bar geometry'' can be turned into ``two-point geometry'' by the addition of two ghost vertices.

All intersections of sticks with the lines $x=L$ and $x=0$ are supposed to be vertices (``entries'' and ``exits'', respectively). 
To apply the above algorithm, we transform a bus-bar geometry into a two-point geometry by adding two ghost vertices, viz., $V_\text{in}$ is adjacent to all the vertices belonging to ``entries'', while  $V_\text{out}$ is adjacent to all the vertices belonging to ``exits''. 
In such a way, the problem of geometrical backbone identification for bus-bar geometry is transformed into one for two-point geometry.

\begin{figure}[!htb]
\includegraphics[width=\columnwidth]{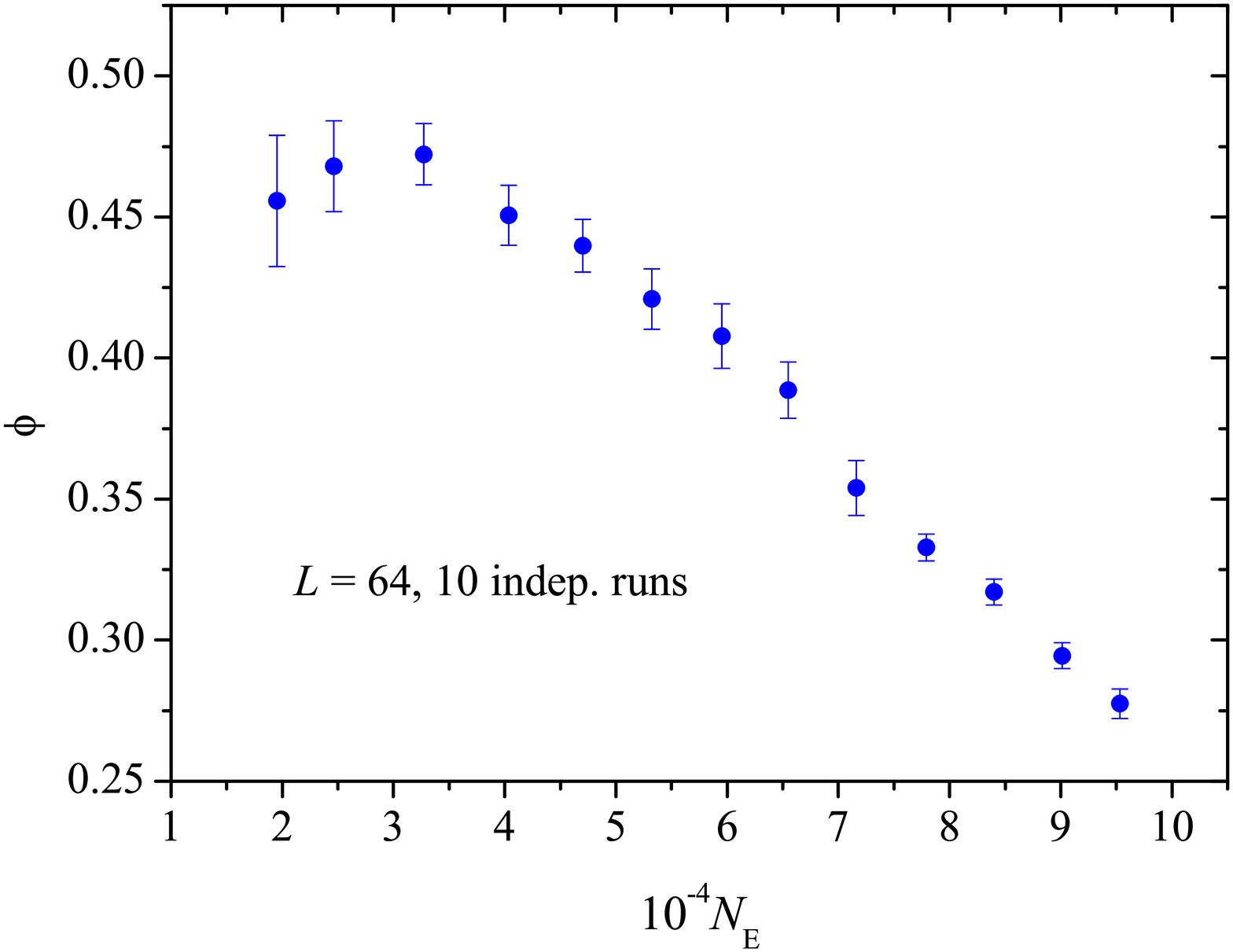}\\
\caption{Fraction of untraversed edges in the ``approximate backbone’’, $\phi$, after complete identification of the backbone against the total number of edges, $N_\text{E}$, in the graph produced by zero-width sticks  whose  centers are i.i.d. within $\mathcal{D}$ ($L=64$).\label{fig:unvisited}}
\end{figure}

For a wide range of number densities, the modified wall follower algorithm can identify the backbone without visiting all the edges of the ``approximate backbone’’. Figure~\ref{fig:unvisited} shows the fraction of untraversed edges in the ``approximate backbone’’ after the complete identification of the backbone, $\phi$, against the total number of edges, $N_\text{E}$, in the graph produced by zero-width sticks, the centers of which are i.i.d. within $\mathcal{D}$. The algorithm is most efficient when the number density of sticks slightly exceeds  the percolation threshold. The fraction of the untraversed edges decreases as the number density increases. Nevertheless, the algorithm is somewhat resource-intensive, since each edge and each vertex have to be stored and tagged, viz., ``untraversed'', ``traversed'', ``backbone''.

To detect the percolation cluster, the Union--Find algorithm~\cite{Newman2000PRL,Newman2001PRE} modified for continuous systems~\cite{Li2009PRE,Mertens2012PRE} was applied. When the percolation  cluster was found, all other clusters were removed since they cannot contribute to the electrical conductivity. All edges of the percolation cluster incident on a vertex of unit valence were cut off, since, obviously, they are simply dead ends. According to Ref.~\onlinecite{Kumar2017JAP}, we denote such a preprocessed percolation cluster as the ``approximate backbone''. To detect the geometrical backbone of the percolation cluster, the modified wall follower algorithm was used (Section~\ref{subsec:MWFA}). When the geometrical backbone has been identified, an adjacency matrix can be formed for it. With this adjacency matrix in hand, Kirchhoff's current law was used for each junction of sticks, and Ohm’s law for each circuit between any two junctions. The obtained set of linear equations with sparse matrix has been solved using~\cite{eigenweb}. Since only square samples were considered, the electrical conductivity is simply the inverse sheet resistance, i.e., $\sigma = R_\Box^{-1}$.

The computer experiments were repeated 100 times for each value of the number density. The error bars in the figures correspond to the standard deviation of the mean. When not shown explicitly, they are of the order of the marker size.

\subsection{Critical behavior and finite-size scaling\label{subsec:scaling}}
The main quantities of interest such as the strengths of both the percolation cluster and its backbone, as well as the electrical conductivity are assumed to demonstrate critical behavior just above the percolation threshold~\cite{Stauffer,Bunde1991percolationI}.

The strength of the percolation cluster, $P_\infty$, varies as
\begin{equation}\label{eq:Pinf}
  P_\infty \propto (n - n_\text{c})^{\beta}, \quad n > n_\text{c},
\end{equation}
where $\beta$ is the critical exponent~\cite{Stauffer}. In 2D, $\beta = 5/36 $~\cite{Stauffer}.
The fraction of bonds in the backbone of the infinite cluster, $P_\text{b}$ varies as
\begin{equation}\label{eq:PBB}
  P_\text{b} \propto (n - n_\text{c})^{\beta_\text{b}}, \quad n > n_\text{c},
\end{equation}
where $\beta_\text{b}$ is the critical exponent~\cite{Shlifer1979JPhysA}. In 2D, $\beta_\text{b}= 1.643\,36(10)$~\cite{Xu2014PRE}.

Near the percolation threshold, $n_\text{c}$, the number of red bonds varies as
\begin{equation}\label{eq:nred}
  n_\text{red} \propto ( n - n_\text{c})^{-1}
\end{equation}
(see Ref.~\onlinecite{Coniglio1982JPA}).

For an insulator--conductor mixture, the electrical conductivity of the mixture, $\sigma$, varies in the vicinity of the percolation threshold, $n_\text{c}$, as
\begin{equation}\label{eq:sigma}
  \sigma \propto \sigma_m (n - n_\text{c})^t, \quad n > n_\text{c},
\end{equation}
where $n$ is the concentration, $\sigma_m$ is the electrical conductivity of the conductor (metal), $t$ is the critical exponent (see, e.g.,~\cite{Clerc1990AdPhys} and the references therein). In 2D, alternative methods give values where the differences slightly exceed the statistical error, e.g., $t = 1.29939(80)$ (the evaluation was performed using the walker diffusion method~\cite{Siclen2018arXiv}) and $t = 1.280(14)$ (Monte Carlo simulations involving the ratio of the stick--stick junction resistance, $R_\text{j}$, to the stick resistance, $R_\text{s}$,~\cite{Li2010PRE}). Estimates of the conductivity  exponent in two dimensions can be found in Ref.~\onlinecite[Table~2]{Hughes2009}.

When a quantity demonstrates critical behavior, it obeys a finite-size scaling (FSS) relationship.
When $\psi$ is a quantity under consideration that obeys an arbitrary power law $\psi \propto (n - n_\text{c})^q$, FSS means that
\begin{equation}\label{eq:scaling}
  \psi = L^{-q/\nu} h\left[L^{1/\nu} (n - n_\text{c})\right],
\end{equation}
where $L$ is the system size, and $h$ is the nonsingular universal scaling function (see, e.g.,~\cite{Hunt2014}), and, in 2D, the critical exponent $\nu = 4/3$~\cite{Stauffer}. 
Accordingly,
\begin{equation}\label{eq:scalingsigma}
   \sigma L^{t/\nu} = h_\sigma \left[L^{1/\nu} \left(n - n_\text{c}\right)\right].
\end{equation}

\section{Results\label{sec:results}}

\begin{figure}[!htb]
\includegraphics[width=\columnwidth]{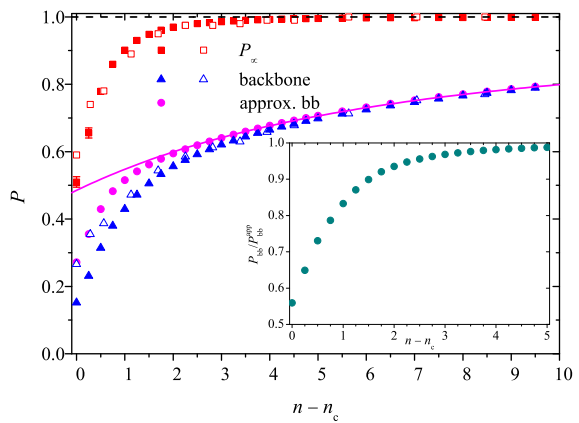}\\
\caption{Dependencies of the strength of the percolation cluster (squares),  of the strength of the backbone (triangles), and of the strength of the ``approximate backbone'' (circles) on the shifted  number density, $n-n_\text{c}$. Solid symbols correspond to our results, while the open symbols reproduce the results extracted from Ref.~\onlinecite{Kumar2017JAP}. The solid curve corresponds to formula~\eqref{eq:betaTDLn}.\label{fig:fractions}}
\end{figure}

Figure~\ref{fig:fractions} demonstrates the dependencies of the quantities of interest on the shifted number density, $n-n_\text{c}$. Solid symbols present our results, while the open symbols reproduce the results extracted from Ref.~\onlinecite{Kumar2017JAP}. The strength of the percolation cluster approaches unity reflecting the fact that almost all sticks belong to the percolation cluster when $n \gtrapprox 2n_\text{c}$. This observation is quite consistent with the previously published results~\cite{Kumar2017JAP}. At the larger number density ($n \gtrapprox 5n_\text{c}$), the backbone and the ``approximate backbone''~\cite{Kumar2017JAP} are indistinguishable within the simulation accuracy. This fact validates  the assumption~\cite{Kumar2017JAP} that, for dense systems, the percolation cluster is identical to its geometrical backbone plus the simplest dead ends, i.e., edges incident on the vertices of unit degree. The solid curve corresponds to a  theoretical estimate of the ``approximate backbone''~\eqref{eq:betaTDLn}. Inset demonstrate the ratio of strength of the backbone to  strength of the ``approximate backbone''. 

\begin{figure*}
\includegraphics[width=\textwidth]{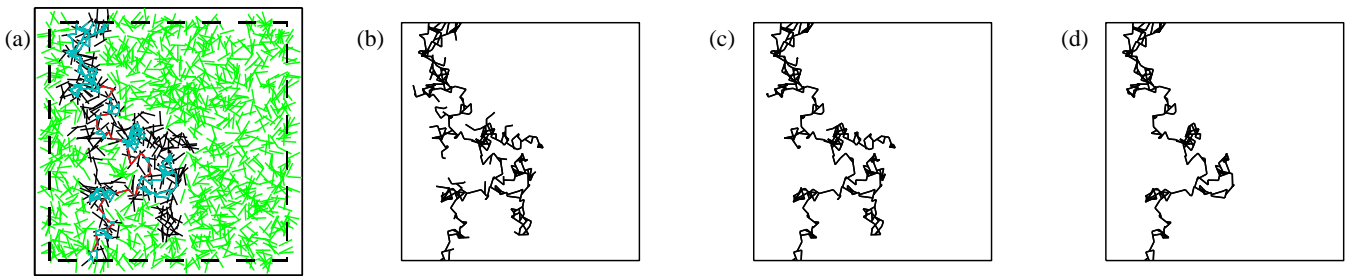}\\
\caption{Example of the system of size $L=16$ under consideration at the percolation threshold. (a)~The percolation cluster, its backbone, dead ends, and red bonds are highlighted. (b)~``Approximate backbone''. (c)~Backbone obtained by iteratively cut off of the edges incident to vertices of unit degree in the percolation cluster as in Ref.~\onlinecite{Kumar2017JAP}. (d)~Backbone.\label{fig:bb}}
\end{figure*}

The strength of the backbone extracted from Ref.~\onlinecite{Kumar2017JAP} (open triangles) is close to our computation of the ``approximate backbone'' when $n \approx n_\text{c}$; then, as the number density increases, it approaches our computation of the backbone. This behavior is explained by the method of backbone identification in Ref.~\onlinecite{Kumar2017JAP}, viz., iteratively cut-off of the edges incident to the vertices of unit degree. In such a way, cycles, which singly connected to the backbone (similar to those presented in Fig.~\ref{fig:bestcase}), are treated as a part of the backbone although they are not. Just above the percolation threshold, such cycles present a significant part of the percolation cluster, although they vanish as the number density increases further [see Fig.~\ref{fig:bb} and compare it with Fig.~1(b), 1(d), and 1(f) in Ref.~\onlinecite{Kumar2017JAP}].

\begin{figure}[!htb]
\includegraphics[width=\columnwidth]{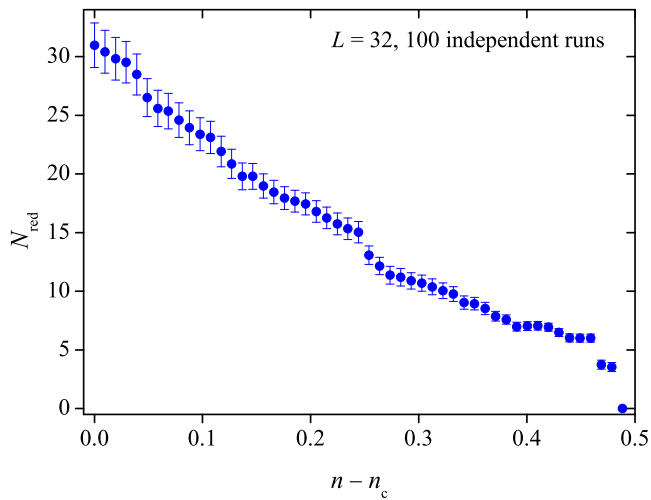}\\
\caption{Dependencies of the number of red bonds on the shifted number density, $n-n_\text{c}$. The results are averaged over 100 independent runs.\label{fig:redbonds}}
\end{figure}

Figure~\ref{fig:redbonds} presents the dependencies of the number of red bonds on the shifted number density, $n-n_\text{c}$. When the number density exceeds the percolation threshold by about 10 per cent, the red bonds vanish. This behavior is not unexpected. When the systems under consideration are dense, red bonds are highly unlikely. The problem of hot spots may be crucial in situations only slightly above the percolation threshold, e.g., in neuromorphic nanowire networks~\cite{OCallaghan2018,DiazAlvarez2019}, whereas it is eliminated in the dense systems suitable for heaters, solar cells and other kinds of transparent electrodes.

\begin{figure}[!htb]
\includegraphics[width=\columnwidth]{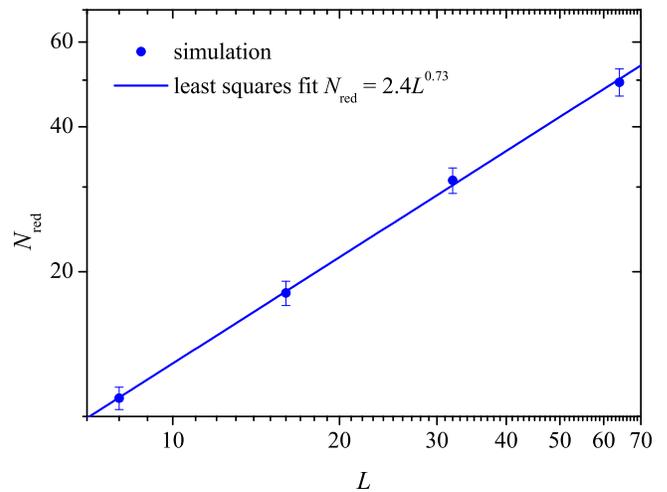}\\
\caption{Number of red bonds at the percolation threshold against the lattice size, $L$, using a log-log scale.\label{fig:rvvsl}}
\end{figure}

Figure~\ref{fig:rvvsl} demonstrates the linear dependency of the number of red bonds in the incipient percolation cluster (i.e., at the percolation threshold) on the system size, using a log-log scale. The least squares fitting leads to $N_\text{red} = 2.4 L^{0.73}$. This behavior is consistent with the theoretical predictions~\cite{Coniglio1982JPA}.

\begin{figure}[!htb]
\includegraphics[width=\columnwidth]{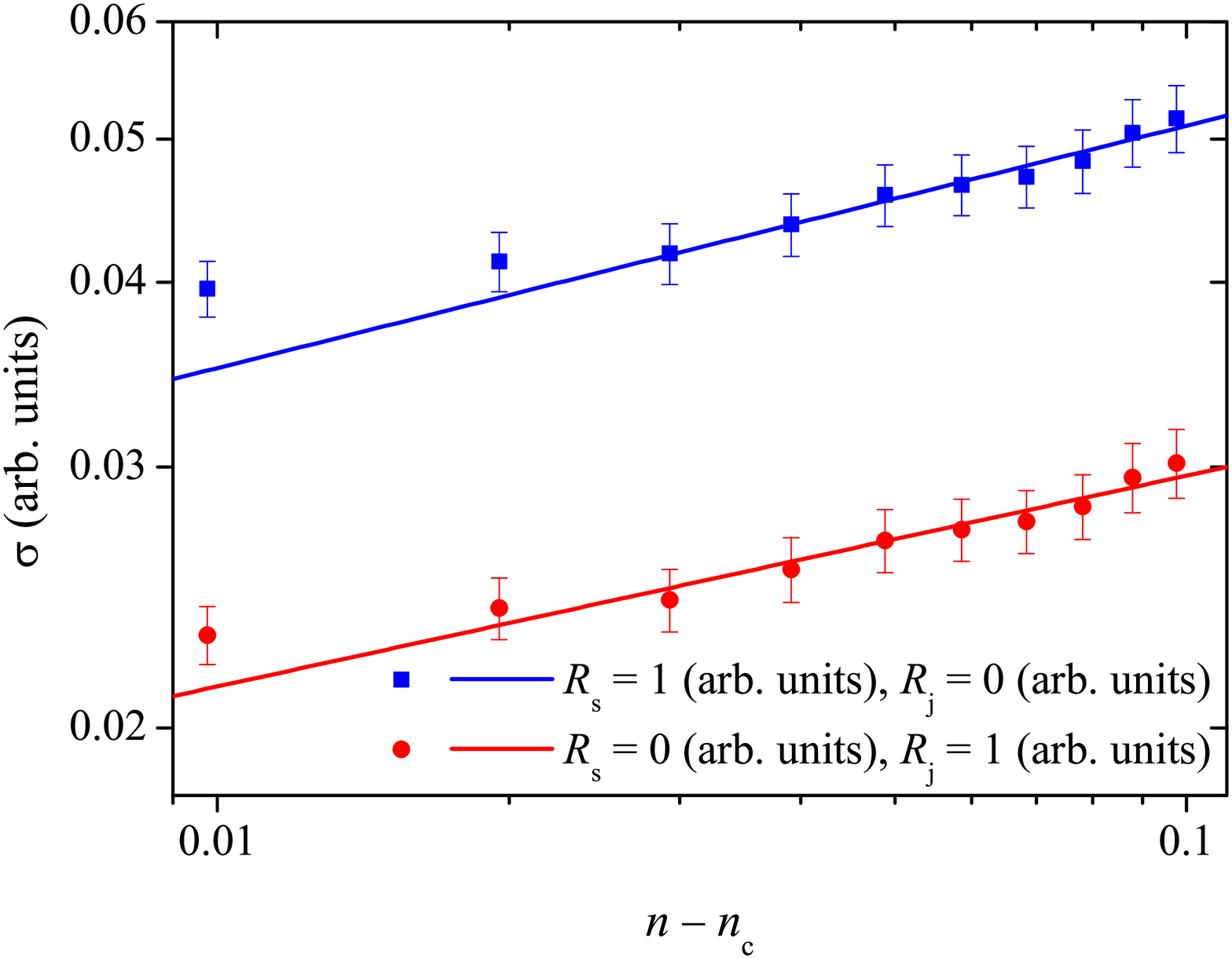}\\
\caption{Behavior of the electrical conductivity near the percolation threshold ($L=32$). The slopes of the lines are $\approx 0.16 \pm 0.01$ in both cases.\label{fig:conductivitynearthreshold}}
\end{figure}
Figure~\ref{fig:conductivitynearthreshold} shows the behavior of the electrical conductivity near the percolation threshold ($L=32$). In the cases of both junction-dominated resistance, $R_\text{j}=1$ (arb. units), $R_\text{s}=0$ (arb. units), and of wire-dominated resistance, $R_\text{j}=0$ (arb. units), $R_\text{s}=1$ (arb. units), an almost linear dependence of the electrical conductivity on the number density can be observed when a log-log scale is used. The linear fit suggests the critical exponent $\approx 0.16 \pm 0.01$ for both cases. This value is significantly smaller than expected.

\begin{figure}[!htb]
\includegraphics[width=\columnwidth]{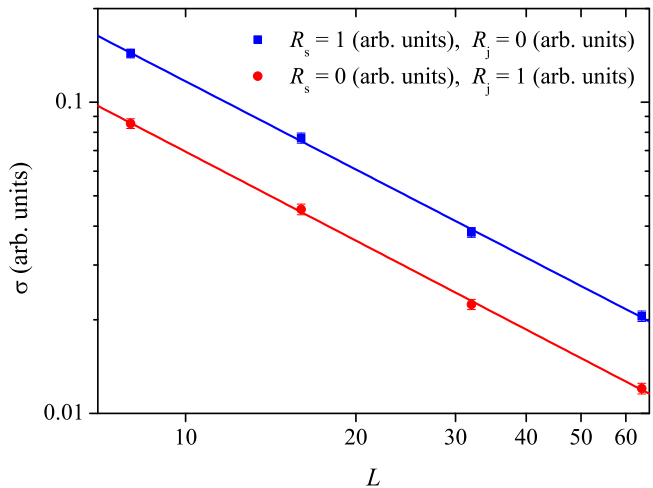}\\
\caption{Electrical conductivity at the percolation threshold, $\sigma$, vs the system size, $L$. The slopes of the lines are $\approx 0.16 \pm 0.01$ in both cases.\label{fig:FSScond}}
\end{figure}

Figure~\ref{fig:FSScond} shows the electrical conductivity at the percolation threshold against the system size. In the cases of both junction-dominated resistance, $R_\text{j}=1$ (arb. units), $R_\text{s}=0$ (arb. units), and of wire-dominated resistance, $R_\text{j}=0$ (arb. units), $R_\text{s}=1$ (arb. units), almost linear dependence of the electrical conductivity on the system size can be observed when a log-log scale is used. The linear fit suggests the critical exponent $t \approx 1.26 \pm 0.02$ for both cases.

\section{Conclusion\label{sec:concl}}

We have considered random resistor networks produced by homogeneous, isotropic, and random deposition of conductive sticks onto an insulating substrate. We have proposed and implemented a modified wall follower algorithm for backbone identification. The algorithm was applied to backbone identification for different number densities of the conductive sticks.
We found that (i) for number densities of sticks above the percolation threshold, the  strength of the percolating cluster quickly approaches unity as the number density of sticks increases; (ii) simultaneously, the percolation cluster is almost identical to its backbone plus the simplest dead ends, i.e., edges that are incident to the vertices of degree one.

Slight above the percolation threshold, the dependence of the electrical conductivity on the concentration of conducting fillers deviates from the expected power law~\eqref{eq:sigma} confirmed by means of computer simulation~\cite{Jagota2015}. This deviation is presumably due to the fact that the volume of the electrically conductive phase is exactly zero at any concentration of fillers, since the model uses zero-width rods. Another possibility may be related to the choice of the number density range that used to find the critical exponent. Since~\eqref{eq:sigma} is hold only near the percolation threshold, the choice of the used range is crucial. Moreover, exponent of the fit is extremely sensitive to the choice of $n_\text{c}$. By contrast, the critical exponent obtained by means of FSS, $t \approx 1.26 \pm 0.02$, is reasonable close to the expected value. 

\acknowledgments
Y.Y.T. and A.V.E. acknowledge the funding from the Foundation for the Advancement of Theoretical Physics and Mathematics ``BASIS'', grant~20-1-1-8-1. The authors would like to thank  I.I.Gordeev for stimulating discussions and A.G.Gorkun for technical assistance.

\bibliography{backboneshort}

\end{document}